\documentclass[smallextended]{svjour3}       
\smartqed  
\usepackage{graphicx}
\usepackage{amsmath}
\usepackage{mathtools}
\usepackage{array}
\usepackage{amsbsy}
\usepackage{subcaption}
\usepackage[numbers]{natbib}
\usepackage{xcolor}
\usepackage[T1]{fontenc}
\usepackage{latexsym}
\usepackage{booktabs}
\usepackage{colortbl}
\usepackage{multicol}
\usepackage{multirow}
\usepackage{rccol}
\usepackage{tabularx}
\usepackage{longtable}
\usepackage{booktabs}
\usepackage{siunitx}
\renewcommand*\thetable{\Roman{table}}
\usepackage{caption}
\usepackage{lipsum}
\usepackage{algorithm}
\usepackage{algpseudocode}
\usepackage{authblk}

\usepackage{comment}
\title{Machine learning adaptation for laminar and turbulent flows: applications to high order discontinuous Galerkin solvers}

\titlerunning{ML adaptation for laminar and turbulent flows}
\authorrunning{Tlales K.}
\author{Kenza Tlales        \and Kheir-Eddine Otmani \and
Gerasimos Ntoukas \and Gonzalo Rubio \and 
Esteban Ferrer 
}

\institute{Kenza Tlales 
, Kheir-Eddine Otmani, Gerasimos Ntoukas (\email{gerasimos.ntoukas@upm.es}), Gonzalo Rubio and Esteban Ferrer   \at
	ETSIAE-UPM - School of Aeronautics, Universidad Polit\'ecnica de Madrid. Plaza Cardenal Cisneros 3, E-28040 Madrid, Spain. //
	Center for Computational Simulation, Universidad Polit\'ecnica de Madrid, Campus de Montegancedo, Boadilla del Monte, 28660, Madrid, Spain. 
	}
\date{Received: date / Accepted: date}

\date{Received: date / Accepted: date}

\begin{document}

\maketitle

\begin{abstract}
We present a machine learning-based mesh refinement technique for steady and unsteady flows. The clustering technique proposed by Otmani et al. \cite{otmaniRobustDetectionViscous2022} is used to mark the viscous and turbulent regions for the flow past a cylinder at Re=40 (steady laminar flow) and Re=3900 (unsteady turbulent flow). Within this clustered region, we increase the polynomial order to show that it is possible to obtain similar levels of accuracy to a uniformly refined mesh. The method is effective as the clustering successfully identifies the two flow regions, a viscous/turbulent dominated region (including the boundary layer and wake) and an inviscid/irrotational region (a potential flow region). The data used within this framework are generated using a high-order discontinuous Galerkin solver, 
allowing to locally refine the polynomial order ($p$-refinement) in each element of the clustered region. For the steady laminar test case we are able to reduce the computational cost up to 32\% and for the unsteady turbulent case up to 33\%.
\end{abstract}



\section{Introduction}
\label{intro}

\noindent 

Within the context of computational fluid dynamics (CFD) and the modeling of complex physical phenomena, there is a continuous effort to reduce the cost of the numerical simulations while increasing the level of accuracy. One methodology along these lines, that has been used extensively across the literature, is to use local refinement (e.g., by augmenting locally to quality of the mesh or by increasing the local polynomial order of approximation within the perspective of higher-order solvers). This allows to reduce the number of degrees of freedom and the associated simulation cost while increasing the solution accuracy, especially in particular regions of interest. This is the case for flows where viscous and turbulent regions can be clearly separated from inviscid regions. Flows belonging to this category include bluff bodies (e.g., cylinders) and aerodynamical shapes (e.g., airfoils). In these types of flow, the boundary layer and wake require local mesh refinement while the potential flow regions, away from walls and the wake, can be resolved using a coarser mesh resolution.

To perform the local refinement it is necessary to define some criteria, which will identify the location 
and the level of refinement required. These sensors can be classified, see~\cite{naddeiComparisonRefinementIndicators2019}, into three distinct categories. The feature-based indicators \cite{ntoukas2021free, ntoukas2022entropy,li2020p}, the local error-based indicators \cite{rueda2019p,de2018flow,basile2022unstructured} and the goal-oriented indicators \cite{laskowski2022functional,dwight2008goal,burgess2011hp}. The feature-based indicators are deduced from physical or theoretical properties of the flow, and by observation of how these influence the different aerodynamical quantities (e.g., lift, drag). These methods are deemed as inexpensive and straightforward to implement, but typically lack robustness and can be case dependent. The second group of indicators, the local error-based, identifies regions that need to be refined depending on high values of quantified error. This can be defined as the difference between the numerical and the exact solution (or an approximation of it). It requires the solution in various levels of meshes and can become relatively expensive before reaching a suitable numerical setup. 
The goal-oriented indicators evaluate the contribution of the error related to a specific target functional (e.g., adjoint-based), but it is costly as a new system of equations needs to be solved. 

In this work, we concentrate on the first approach, the feature-based adaptation, since it constitutes a cost-effective method of locating the regions of interest and adjusting the local degrees of freedom. We propose a new robust methodology which removes some of the typical drawbacks of feature-based methods. For instance, one of the drawbacks of feature-based adaptation is the use of sensors which are related to physical quantities. An arbitrary threshold parameter is used to define the isosurface based on the feature.  The isosurface defines the boundary between the regions requiring refinement and the rest of the domain. The threshold defining the isosurface is an arbitrary user-defined parameter that defines the region (or elements in the mesh) selected for refinement. Feature based adaptation typically needs an iterative trial-and-error process to define a suitable threshold parameter and to obtain convergence in the aerodynamic functionals (e.g., lift, drag). This paper employs a machine learning-based methodology to select the refinement regions while avoiding the iterative process needed to specify it.

The development of machine learning (ML) techniques during the past decade has percolated into natural sciences and engineering \cite{brunton_kutz_2019,GARNIER2021104973,vinuesa2022enhancing}. Specifically, on the topic of flow region recognition, ML and CFD have been combined using a supervised machine learning-based framework~\cite{liUsingMachineLearning2020} to identify the flow state (turbulent or non-turbulent flow). Different training datasets at different Reynolds numbers are used to train Extreme Gradient Boosting (XGBoost). This model can efficiently generate a robust detector for identifying turbulent regions for a flow past a circular cylinder. In the same context, a machine learning algorithm has been used in \cite{saettaIdentificationFlowFielda} to distinguish the grid cells that contain part of the boundary layer, shock waves, or external inviscid flow from RANS solutions. 
Within the framework of mesh adaptation, The authors of \cite{fidkowskiMetricbasedGoalorientedMesh2021} introduced artificial neural networks to perform $h$-adaptation by predicting the desired element aspect ratio from readily accessible features of the primal and adjoint solutions. The neural network models not only reproduced the correct anisotropy for cases in the training data set, but they could also generalize to flow fields not considered during training.
Yang et al. \cite{yangReinforcementLearningAdaptive2021} proposed a work in $h$-refinement in which a feedforward neural network is used for adaptive mesh refinement.

In this work, we study the capabilities of the clustering technique developed in \cite{otmaniRobustDetectionViscous2022} as a mesh adaptation sensor. This clustering technique is capable of separating the flow into two regions, a viscous and turbulent dominated region (boundary layer and wake region) and an inviscid, irrotational region (outer flow region). The clustering technique is effective, as presented in the numerical experiments of this paper, in turbulent and laminar test cases. The Gaussian Mixture Model (GMM) model along with a feature space used in \cite{otmaniRobustDetectionViscous2022} is parameter free and does not require the tuning of any threshold value.

The viscous-dominated region requires a finer mesh than the inviscid region. Therefore, we propose to use a high-order polynomial approximation to solve the viscous-dominated region and a low-order polynomial approximation for the inviscid region. The aim is to accurately predict the flow features and get an equivalent solution to that of a uniform high polynomial order at a reduced cost. One advantage of having only two regions is that the computational overhead related to the mortar computation is reduced (compared to other techniques that use different polynomial orders for every element in the domain (\cite{rueda2019p, kompenhans2016adaptation, kompenhansComparisonsPadaptationStrategies2016}). 
Additionally, the clustering-adaptation is performed only once when the flow is developed (for steady and unsteady simulations). This simplifies the load-balancing process required for parallel computations. 

The rest of this paper is organized as follows. Initially, we present the methodology in section \ref{sec:meth}, including the clustering technique and the adaptation algorithm. Then, in section \ref{sec:results} we present the results along with the analysis of the numerical experiments for the flow past a cylinder at Re=40 (steady laminar regime) and Re=3900 (unsteady turbulent regime) .


\section{Methodology}\label{sec:meth}
\noindent In this section, we define the clustering technique and introduce the details of the flow invariants necessary for the cluster. Also, we present the adaptation algorithm and a brief description of the high-order solver used to carry out the simulations.


\subsection{Clustering of viscous/turbulent  flow region}

\noindent To classify different flow regions, we use the technique proposed in \cite{otmaniRobustDetectionViscous2022}, where the principal invariants of the strain and rotational rate tensors were used as input to the Gaussian Mixture Model(see nest section) to identify the two flow regions (a viscous rotational and inviscid irrotational region).
The feature space used as input to the GMM relies on Galilean invariants of the flow. The feature space is ${E=\left\{Q_S,R_S,Q_{\Omega}\right\}}$, where $Q_S$ and $R_S$ are the two principal invariants of the strain rate tensor $S$ defined as 
\begin{align}
    Q_{S}&=\frac{1}{2}\left((\mathrm{tr}(S))^2-\mathrm{tr}(S^2)\right),\\
    R_{S}&=-\frac{1}{3}\mathrm{det}(S).
\end{align}
$S$ is the strain rate tensor given by $S=(J+J^T)/2$ and $J=\nabla U$ is the gradient of the velocity field $U=(u,v,w)$.
The first invariant of the strain rate tensor $Q_S$ is relevant in indicating regions of high local viscous dissipation rate of the kinetic energy $\epsilon = -4\mu Q_S$ \cite{Zhou2015OnTE}, where $\mu$ is the dynamic viscosity of the flow. $R_S$ is the second strain rate tensor invariant which indicates strain production or destruction in the flow field
\cite{da_silva_2008}.
%
The rotational rate tensor $\Omega$ has one invariant defined as:
\begin{equation}
Q_\Omega=-\frac{1}{2}\left(\mathrm{tr}(\Omega^2)\right),
\end{equation}
where $\Omega=\left(J-J^T)\right/2$. $Q_\Omega$ indicates rotational regions in the flow field as it is proportional to enstrophy density \textbf{$\xi$} \cite{Zhou2015OnTE}. 
The GMM along with the feature space $E$ is used to detect two flow regions, a rotational viscous dominated region and an inviscid irrotational region. This methodology is applied in two different flow regimes, a laminar flow past a cylinder at $Re=40$ and a turbulent flow past a cylinder at $Re=3900$.

\subsection{Gaussian Mixture Model(GMM)}
The Gaussian Mixture Model is an unsupervised machine learning algorithm that assigns data to different cluster membership probabilities. Each point in the data will follow a certain Gaussian distribution based on a probability measure. The number of clusters must be selected before training.
\par
The GMM considers the data as a combination of normal distributions. Each of the normal distributions is characterized by its mean and variance parameters. This model estimates these parameters by performing a maximum likelihood estimation. One of the common methods used to perform this estimate is the expectation-maximization method \cite{em_method}. The expectation maximization method is an iterative process to estimate the parameters of a probability density through maximization of a likelihood function. This process is essentially done in two steps, the expectation step and the maximization step. In the expectation step, the conditional expectation of the given data is computed after randomly initializing the parameters. In the maximization step, the likelihood of the data is maximized and then a new estimation of the parameters is given. The algorithm will repeat the expectation and maximization steps iteratively until it reaches the convergence tolerance where the expected values computed in the expectation step do not change significantly.
The number of normal components (clusters) is $N=2$ as we want to identify two distinct flow regions, a viscous/turbulent-dominated rotational region (boundary layer and wake region) and an inviscid irrotational region (outer region). This model is implemented using the Gaussian mixture class in the scikit-learn Python library \cite{pedregosaScikitlearnMachineLearning2011}.

\subsection{Cluster based adaptation}
The regions identified by the clustering technique will be used for local $p$-adaptation. The clustering technique will assign to each degree of freedom in the mesh two membership probabilities of belonging to the regions of interest. The probability $p_c$ for the clustered viscous/turbulent region and $p_i$ for the outer region. Within each element, the mean of membership probabilities ${p_c}_j\big|_{j=1}^N$, ${p_i}_j\big|_{j=1}^N$ for each region is computed as follows,
\begin{equation}
  \bar{p}_v= \frac{1}{N}\sum_{j=1}^N {p_c}_j, \qquad {p_c}_j \in [0,1],
\end{equation}
\begin{equation}
  \bar{p}_i= \frac{1}{N}\sum_{j=1}^N {p_i}_j, \qquad {p_i}_j \in [0,1],
\end{equation}
where $N=(P+1)^{3}$ is the number of degrees of freedom of each element of the mesh, and $P$ denotes the polynomial order associated to the high order discretization. An element will be assigned to the region with the highest mean of membership probabilities $\bar{p}=\max{(\bar{p}_i,\bar{p}_v)}$. The same process will be applied for all elements to supply each one with a region ID, viscous/turbulent or inviscid. Having marked the elements belonging to each region, we can now modify the polynomial order in each region accordingly.\\

The cluster based adaptation involves three steps. In the first step we run a simulation with a polynomial order, $P=P_{cluster}$, until the flow is fully developed. $P_{cluster}$ is chosen so that the solver captures, with the desired level of accuracy, all the features of the flow. 
In the second step we perform the clustering, as explained in the previous paragraph.
In the third step, we use the ID of each element, depending on which region it belongs to, and assign the corresponding polynomial order within that element. In the present work, we coarsen the polynomial in the inviscid regions, $P_{inviscid}$, while keeping the original high-order polynomial, $P_{cluster}$, in the elements belonging to the viscous/turbulent region. It is reasonable to keep $P_{cluster}$ in the viscous region, as it was selected to achieve a certain level of accuracy. However this high resolution can be relaxed using $P_{inviscid}<P_{cluster}$ in the inviscid region, that presents a lower flow complexity. 
However, different strategies to perform adaptation based on this clustering may be proposed (e.g., $h$-refinement/coarsening).
Finally, the simulation is restarted with the adapted polynomial order from the same instant until we reach the final solution. This process is summarized in algorithm~\ref{alg:cap}.

 \begin{algorithm}
\caption{Methodology algorithm}\label{alg:cap}
\begin{algorithmic}
     \State Start the simulation using uniform polynomial order $P_{cluster}$
     \State Stop the simulation when the flow is fully developed
     \State Train GMM with feature space $E$
     \State Return $p_c$ and $p_i$ \Comment{Membership probabilities for every point}
     \For{$k$ in number of elements}
         \State $\bar{p}_i = \frac{1}{N}\sum_{j=1}^N {p_i}_j$
         \State $\bar{p}_v = \frac{1}{N}\sum_{j=1}^N {p_c}_j$
         \If{$\bar{p}_v \geq \bar{p}_i$}
             \State $P(k) \gets P_{cluster}$ \Comment{high-order polynomial in the viscous rotational region}
         \ElsIf{$\bar{p}_v < \bar{p}_i$}
             \State $P(k) \gets P_{inviscid}$ \Comment{low-order polynomial in the inviscid irrotational region}
         \EndIf
     \EndFor
     \State Return $P$
     \State Restart the simulation using $P$
\end{algorithmic}
\end{algorithm}


\subsection{High order discontinuous Galerkin solver} 
The data used in this work have been generated using the high-order spectral element CFD solver HORSES3D \cite{HORSES3D}. HORSES3D is a 3D parallel code developed at ETSIAE–UPM (the School of Aeronautics of the Polytechnic University of Madrid).
This framework uses the high-order discontinuous Galerkin spectral element method (DGSEM) and is written in modern Fortran 2003. It targets simulations of fluid-flow phenomena such as those governed by compressible and incompressible Navier-Stokes equations, and it supports curvilinear hexahedral meshes of arbitrary order. There are several options of subgrid turbulence models for the users to choose from for LES simulations such as the Smagorinsky subgrid model, Wale and Vreman. One of the main features offered in HORSES3D is its support for anisotropic $p$-non-conforming elements, as the polynomial order can be varied independently in each element in each direction. In this work, we make use of the adaptation capability to increase/decrease the polynomial order uniformly within each element in the regions determined by the clustering technique.


\section{Results and Discussion}\label{sec:results}
The mesh refinement strategy is applied to a steady flow test case around a cylinder at Re=40 and an unsteady turbulent flow around a circular cylinder at Re=3900, with a Smagorinsky subgrid closure model.

\subsection{Flow past a cylinder at Re=40} 
\begin{figure*}[!htb]
    	\begin{subfigure}[b]{0.5\linewidth}
    		\centering
    		\includegraphics[width=0.90\linewidth,height=5.00cm,keepaspectratio]{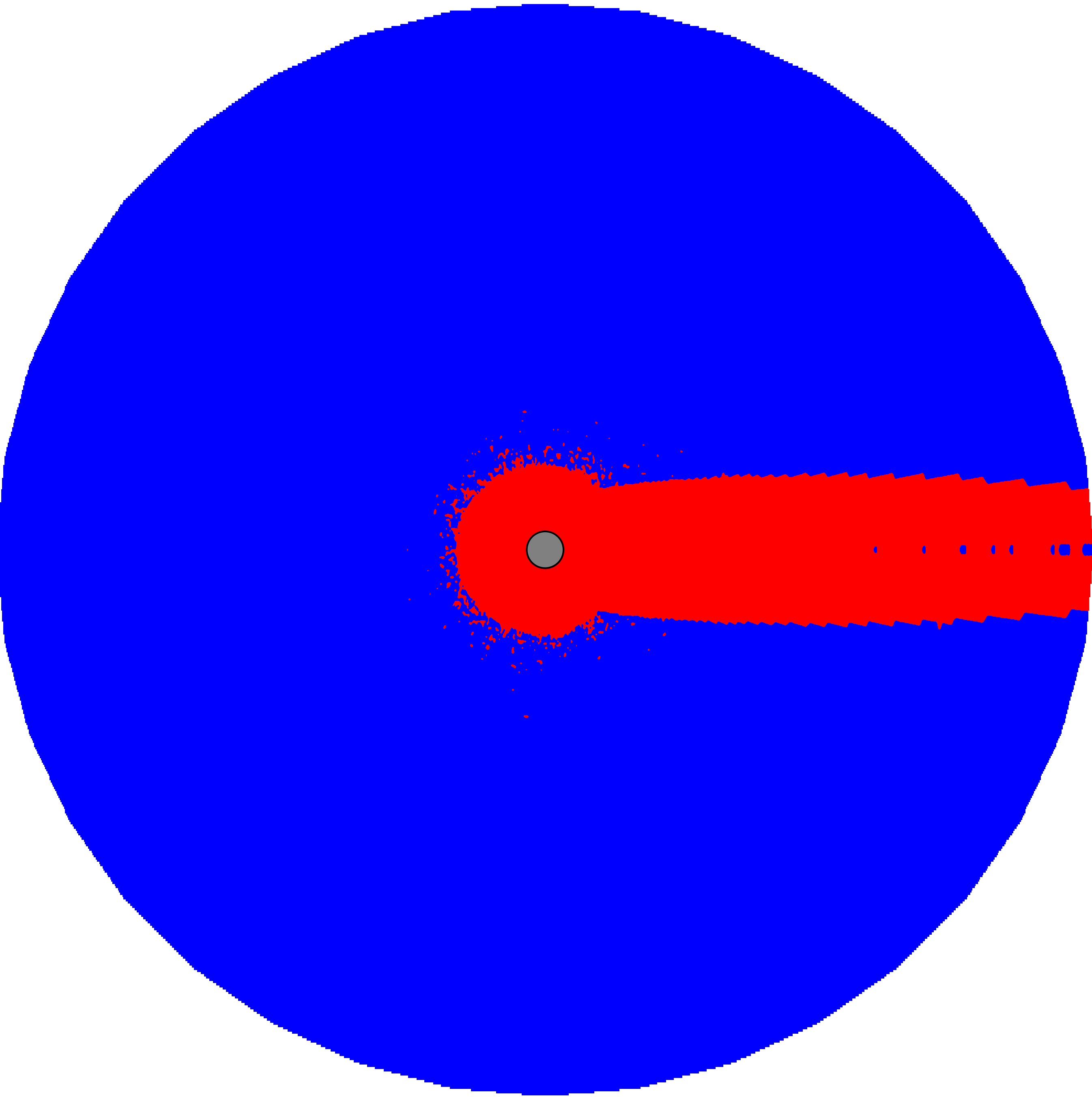}
    \caption{}
    		\label{Fig:reg40}
    	\end{subfigure}%
    	\begin{subfigure}[b]{0.5\linewidth}\qquad
    		\includegraphics[width=0.90\linewidth,height=5.00cm,keepaspectratio]{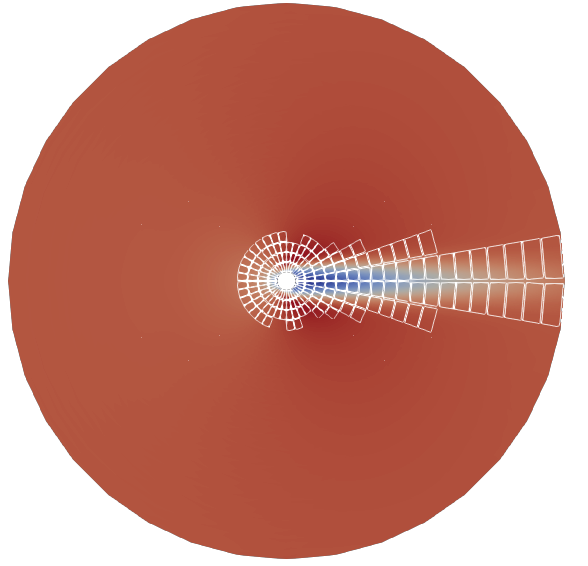}
    \caption{}
    		\label{Fig:ref40}
    	\end{subfigure}
    	\caption{Regions detected with the GMM along with the feature space $E$ for flow past a cylinder at $Re=40$. \textcolor{red}{Red}: viscous rotational region. \textcolor{blue}{Blue}: outer region (\ref{Fig:reg40}), selected elements for high-order polynomial refinement (\ref{Fig:ref40}) for the flow past a circular cylinder at $Re=40$}
    \end{figure*}

In this test case, we refine the mesh for steady laminar flow around a cylinder at Re=40.
The steady flow data are generated in a mesh of 684 elements with 85500 degrees of freedom using a uniform polynomial order $P=4$. 
The adaptation is carried out following the procedure described in algorithm~(\ref{alg:cap}). Figure (\ref{Fig:reg40}) shows the regions identified using the clustering GMM model. The initial, high polynomial order is kept in the selected elements in figure (\ref{Fig:ref40}) based on the mean of probabilities to belong to the viscous rotational region. For this numerical experiment we use $P_{cluster}=4$ for the viscous regions and $P_{inviscid} \in [1,2,3]$ for the inviscid ones.

\begin{figure*}[!h]
	    \centering
			\includegraphics[width=0.90\linewidth,height=5.00cm,keepaspectratio]{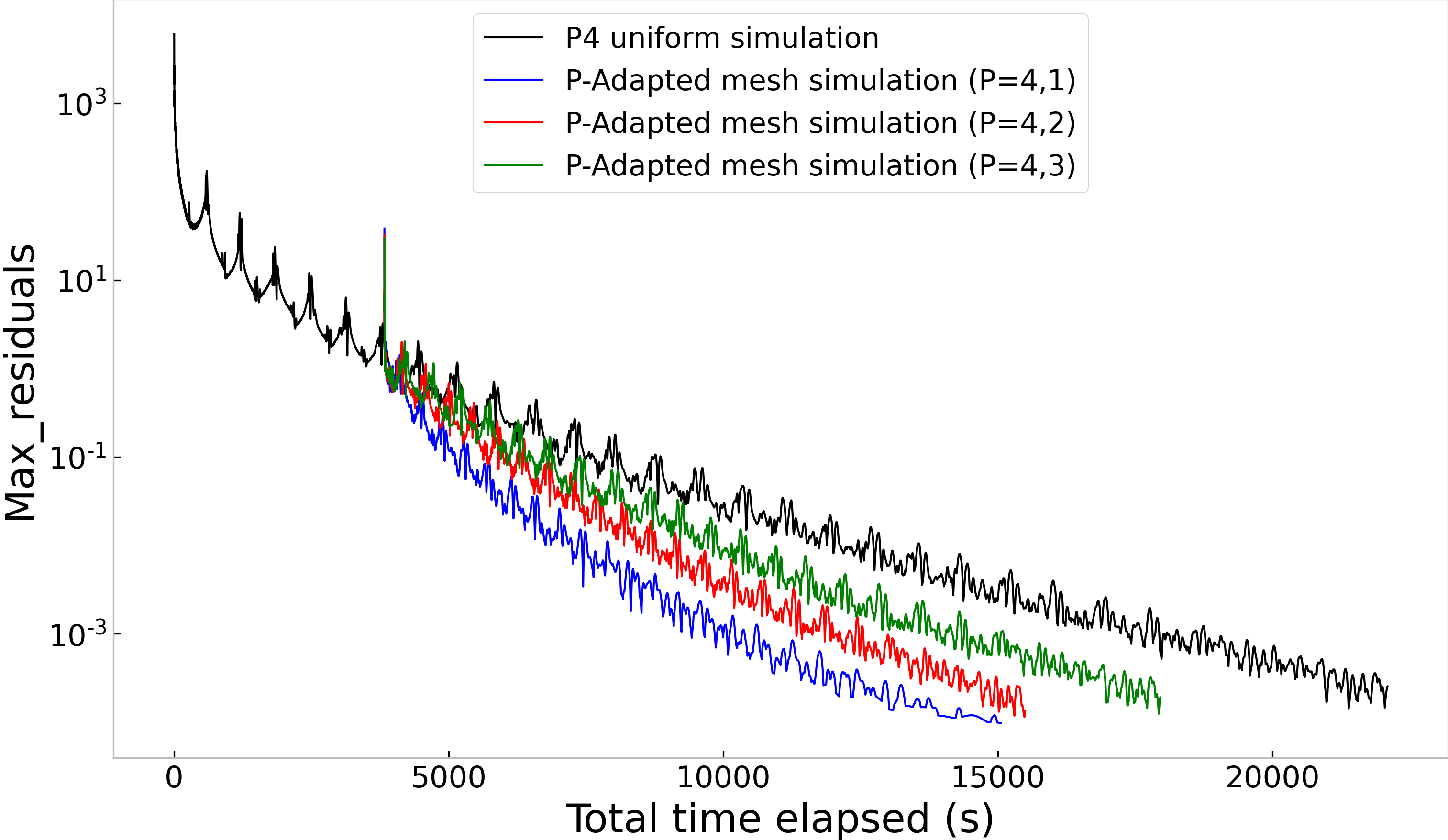}
		\caption{Max residuals for homogeneous and $P$-adapted solutions}
		\label{Fig:convergence}
	\end{figure*}

We run all simulations until the maximum residuals decreases by a factor of 1000. At this point the flow is not converged yet, but the wake has developed. We can use this data to perform the clustering and detect the regions that need adaptation/coarsening. Having adapted, we converge the simulation until the maximum residual is $10^{-4}$.
This adapt and converge technique reduces significantly the computational time. Figure \ref{Fig:convergence} shows the convergence history of the homogeneous $P=4$ solution and the clustering adapted solution ($P_{cluster}=4$ and $P_{inviscid}\in [1,2,3]$), where it can be observed that convergence is accelerated when adaptation is performed. The computational savings and the corresponding degrees of freedom for each simulation are presented in table~\ref{tab:tab2}. 
The adapted solutions for $P_{inviscid} \in [1,2,3]$ exhibit $32\%, 29\% \text{ and } 19\%$ reduction of computational time accordingly compared to the homogeneous $(P = 4)$ solution and $67\%, 55\% \text{ and }$ $35\%$ less degrees of freedom.

\begin{table}[h]                           
	\centering
	\caption{Results shown for residuals $=10^{-4}$ (converged solution).}
	\label{tab:tab2}
	\renewcommand{\arraystretch}{1.25}
	\begin{tabular}{cccccc}
		\toprule
		$P_{\mathrm{cluster}}$ & $P_{\mathrm{inviscid}}$ & Adapted & computational time(s) & $\mathrm{C}_{d}$ & DoF     \\
		\toprule
		$2$ & $2$ & No  & $\num{8.39E+03}$ & $\num{1.5062}$ & $\num{18468}$ \\
		$3$ & $3$ & No  & $\num{1.23E+04}$ & $\num{1.5264}$ & $\num{43776}$ \\
		$4$ & $4$ & No  & $\num{2.21E+04}$ & $\num{1.5221}$ & $\num{85500}$ \\
		$4$ & $1$ & Yes & $\num{1.50E+04}$ & $\num{1.5214}$ & $\num{27936}$ \\
		$4$ & $2$ & Yes & $\num{1.58E+04}$ & $\num{1.5225}$ & $\num{38460}$ \\
		$4$ & $3$ & Yes & $\num{1.79E+04}$ & $\num{1.5222}$ & $\num{55488}$ \\ 
		\bottomrule
	\end{tabular}
\end{table}

Finally, table~\ref{tab:tab1} reports drag values extracted from the literature. Comparing tables~\ref{tab:tab1} and \ref{tab:tab2}, we observe that the baseline homogeneous (for P=4) and all clustered adapted solutions ($P_{cluster}=4$ and $P_{inviscid}\in [1,2,3]$) provide accurate predictions for the drag coefficient.

\begin{table}[h]                           
	\centering
	\caption{Comparison of the results for the drag coefficient ($C_d$ ) for flow past a cylinder at Re=40}
	\label{tab:tab1}
	\renewcommand{\arraystretch}{1.25}
	\begin{tabular}{cc}
		\toprule
		Case & $C_d$  \\
		\toprule
		Dennis and Chang \cite{dennis1970numerical} & $1.52$ \\
		Fornberg \cite{fornberg1980numerical}       & $1.50$ \\
		Choi et al. \cite{CHOI2007757}              & $1.49$ \\
		\bottomrule
	\end{tabular}
\end{table}

This first laminar steady case shows that the GMM clustering can separate the viscous region (the boundary layer and wake) from the outer inviscid. Refinement of the clustered region and coarsening of the inviscid outer region shows that the detected clustered region is indeed responsible for errors in drag, and that refining it increases the accuracy of the simulations. In addition, we conclude that the clustering adaptation is an efficient technique for mesh adaption in steady flows and can help to speed up convergence.

\subsection{Flow past a cylinder at Re=3900}
\begin{figure*}[!h]
    	\begin{subfigure}[h!]{0.4\linewidth}
    		\centering
			\includegraphics[width=0.9\linewidth]{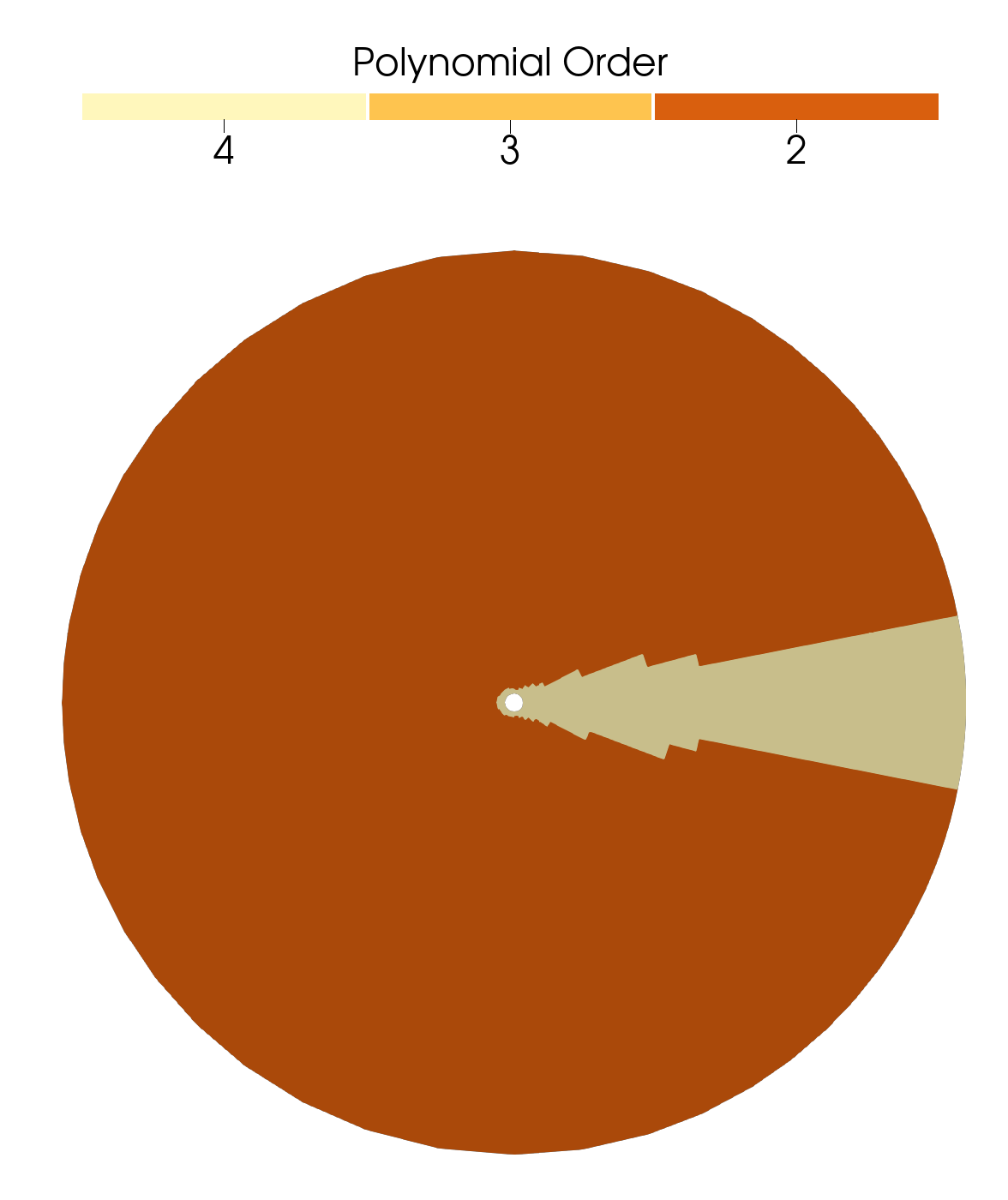}
    \caption{}
    		\label{fig:cyl_adapted}
    	\end{subfigure}%
    	     \hfill
    	\begin{subfigure}[h!]{0.6\linewidth}
			\includegraphics[width=1.0\linewidth]{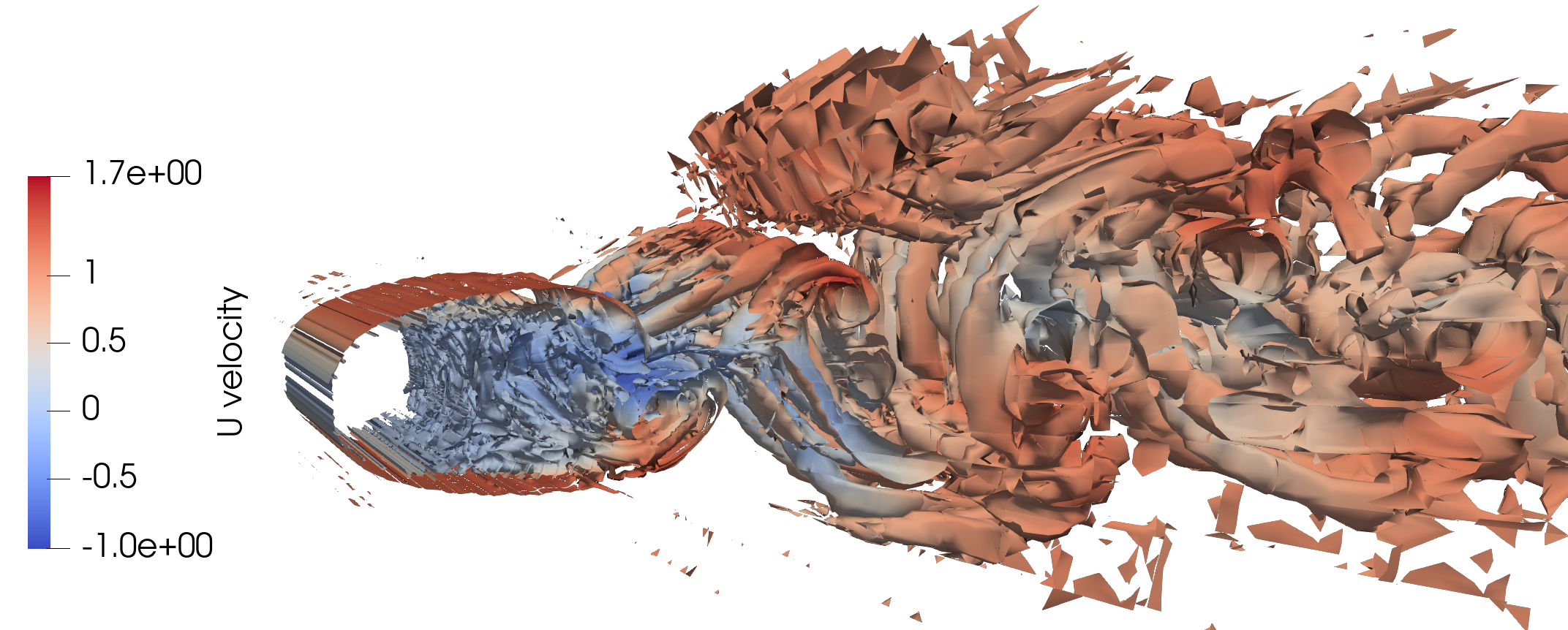}
    \caption{}
    		\label{fig:cyl_contour}
    	\end{subfigure}
    	\caption{Polynomial order distribution from the clustering algorithm for the cylinder at Re=3900 test case presented in \ref{fig:cyl_adapted}. \ref{fig:cyl_contour} presents an isosurface  of the $\mathrm{Q}_{crit}=0.1$ coloured with the axial velocity $u$}
    \end{figure*}

This case has been considered extensively by numerous researchers within the numerical \cite{parnaudeau2008experimental,ma2000dynamics} and experimental spectrum \cite{parnaudeau2008experimental,norbergeffects,lourenco1994characteristics}. For this case we use a $2^{nd}$ order mesh of 20736 hexahedral elements. The mesh has been extruded in the spanwise direction as $L_z/D=\pi$ and subdivided into 16 elements along this direction. The solution has been calculated using a uniform polynomial order of $P\in{[3,4]}$, as well as an adapted solution defined with the clustering algorithm with $P_{cluster}=4$ and $P_{inviscid}=2$. In this turbulent case, using $P_{inviscid}=1$ resulted in diverging simulations probably due to aliasing instabilities at the interface between elements  with different polynomial orders (mortar surfaces), for this reason  we only report results for $P_{inviscid}=2$. The adaptation has been performed using the clustering algorithm on a single instant of the flow field after the wake has developed. The adapted mesh is presented in figure \ref{fig:cyl_adapted}. A constant non-dimensional time step of $\Delta t = 2 \times 10^{-3}$ has been used throughout all the different simulations. To minimize the aliasing errors and ensure that the method is robust, a split form discretization has been employed with Pirozzoli averaging \citep{pirozzoli2010generalized}.

To measure statistical quantities, we follow \cite{parnaudeau2008experimental} and allow the flow to develop for a time interval of $t=150 D/U_{c}$ to remove the influence of the initial condition. Then the statistics are gathered for a time duration of approximately 100 shedding cycles for each of the simulations carried out, which is in line with the averaging intervals in the literature \cite{parnaudeau2008experimental}.

The results for the mean drag coefficient $\mathrm{C}_{d}$, Strouhal number $\mathrm{St}$ and the recirculation length $\mathrm{L}_{r}$ are presented in table~\ref{tab:cyl3900_tab}. We compare the results obtained from HORSES3D with uniform $P=3$, $P=4$ solutions and a p-adapted mesh against numerical and experimental results from previous works. As presented, the uniform $P=4$ and the adapted results fall well within what is reported in the literature. For the uniform $P=3$ solution, we observe that the mean drag value is severely under-predicted due to the lack of sufficient resolution, even though the Strouhal number is close to the reported values.

\begin{table}[!htb]
\centering
\caption{Comparison of the mean statistical quantities for the cylinder at Re = 3900 HORSES3D and the literature. We compare the Strouhal number $\mathrm{St}$, the drag coefficient $\mathrm{C}_{d}$ and the recirculation length $\mathrm{L}_{r}$. All simulations used a spanwise length $\mathrm{L}_{z}/D=\pi$ }
\begin{tabular*}{0.7\textwidth}{c @{\extracolsep{\fill}} |cccc}
\hline
   & $\mathrm{St}$ & $\mathrm{C}_{d}$ & $\mathrm{L}_{r}$ & $\mathrm{L}_{z}\backslash	D$  \\ \hline
Uniform P3     & 0.202   & 0.7844 & 1.36 & $\pi$ \\ 
Uniform P4      & 0.203   & 0.9513 & 1.64 & $\pi$ \\ 
Cluster-Adapt P4-P2         & 0.204   & 0.9506 & 1.63 & $\pi$ \\ \hline
Parnadeau et al.\cite{parnaudeau2008experimental}     & 0.208   &   -    & 1.56 &  $\pi$      \\ 
Snyder and Degrez \cite{snyder2003large}          & 0.207   & 1.09 &  1.30   & $\pi$  \\ 
Kravchenko and Moin\cite{kravchenko2000numerical}        & 0.210   & 1.04 &  1.35   & $\pi$\\ 
Breuer \cite{breuer1998large}         & -   & 1.07 &  1.20   & $\pi$ \\ 
Franke and Frank \cite{franke2002large}         & 0.209   & 0.98 &  1.64   & $\pi$  \\ 
(DNS) Ma et al. \cite{ma2000dynamics}      & 0.219   & 1.59 &  -   & $\pi$\\ 
Ouvrard et al. \cite{highordermethouvrard2010classical}      & 0.223   & 0.94 &  1.56   & $\pi$\\ \hline
\end{tabular*}
\label{tab:cyl3900_tab}
\end{table}

\begin{figure}[!htb]
\centering
\subfloat[ $<u',u'>$]{\label{fig:uu}\includegraphics[width=.45\linewidth]{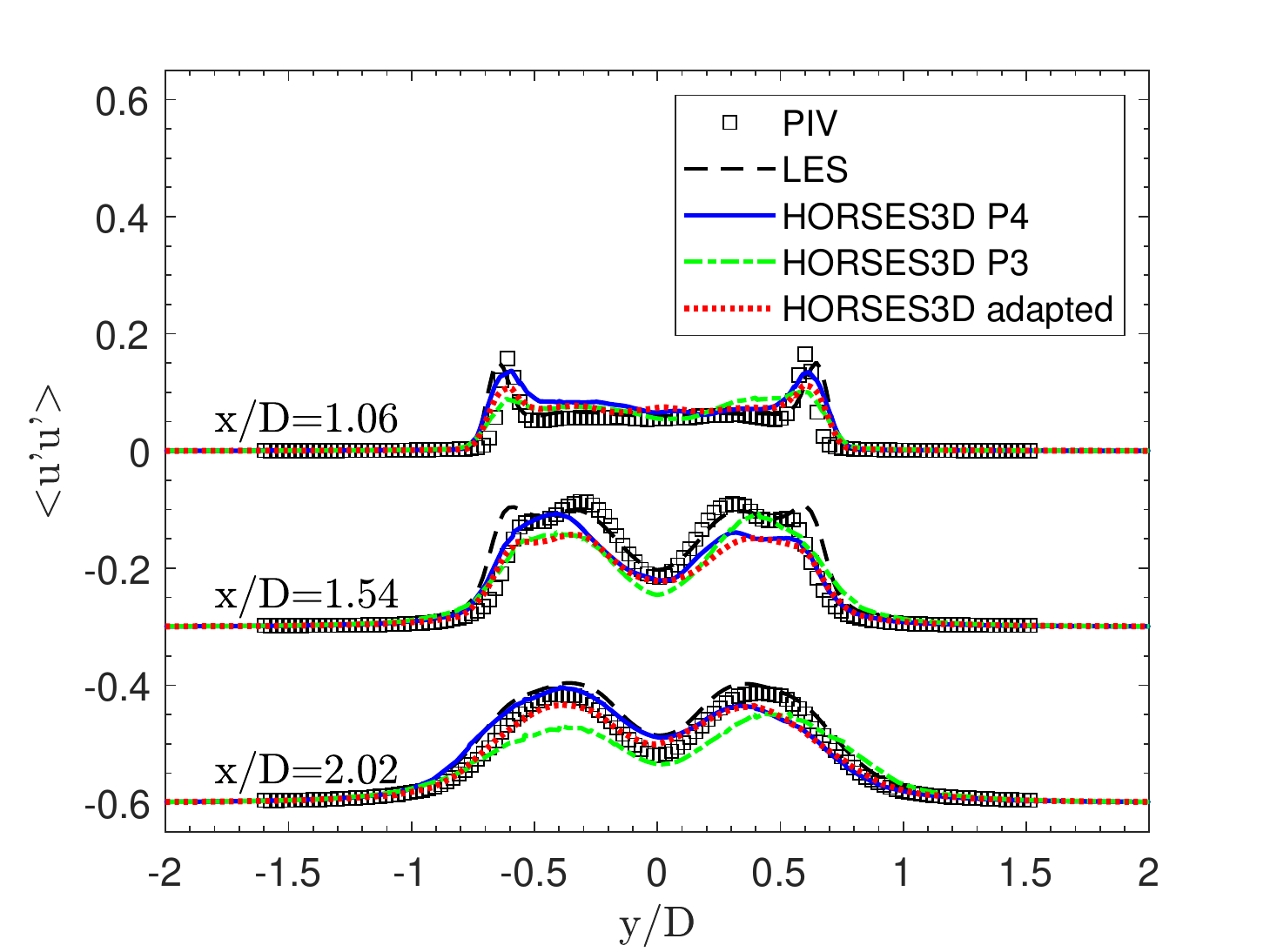}}
\subfloat[$<u',v'>$]{\label{fig:uv}\includegraphics[width=.45\linewidth]{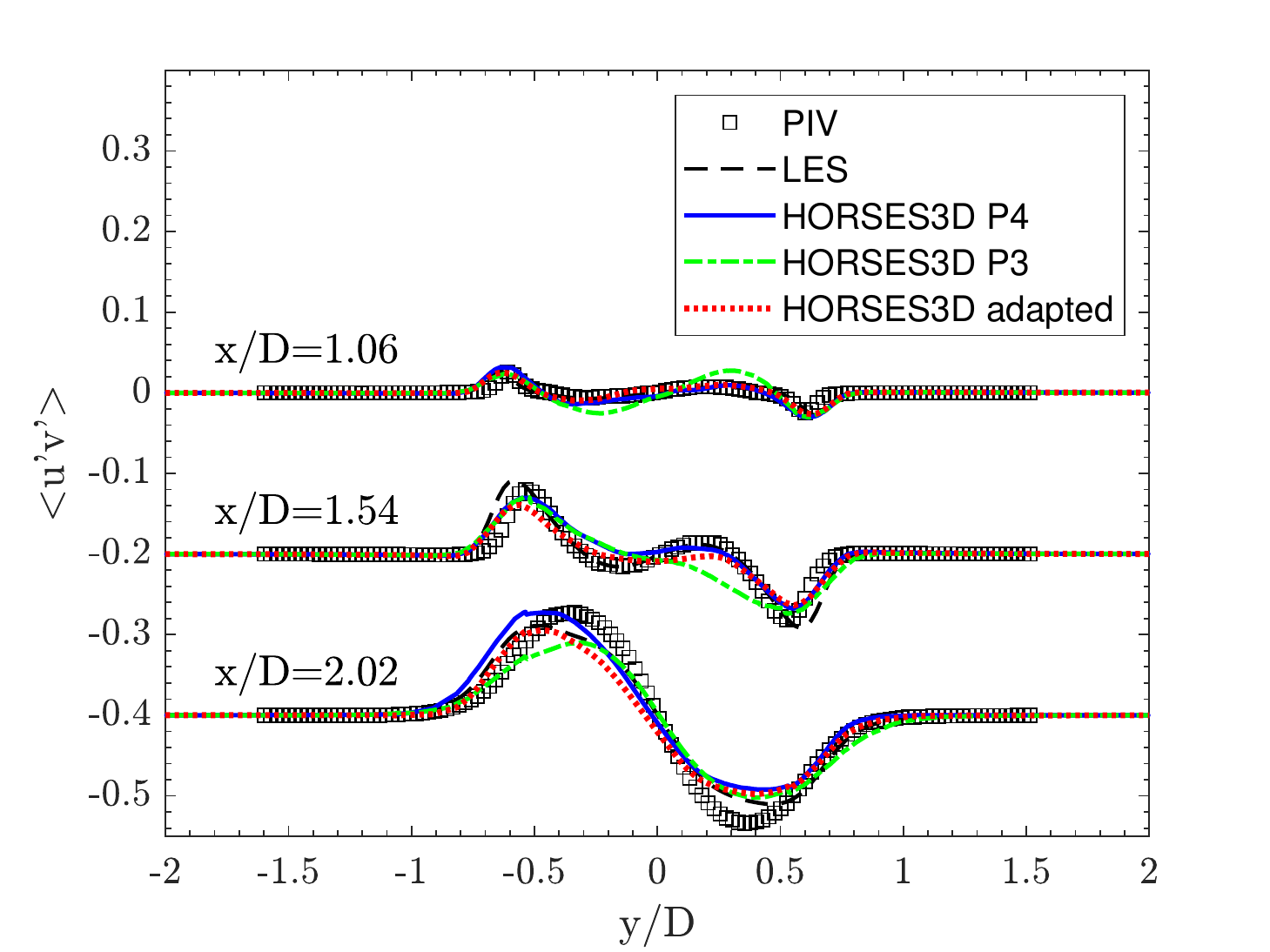}}\par 
\subfloat[$U_{mean}$]{\label{fig:uc}\includegraphics[width=.45\linewidth]{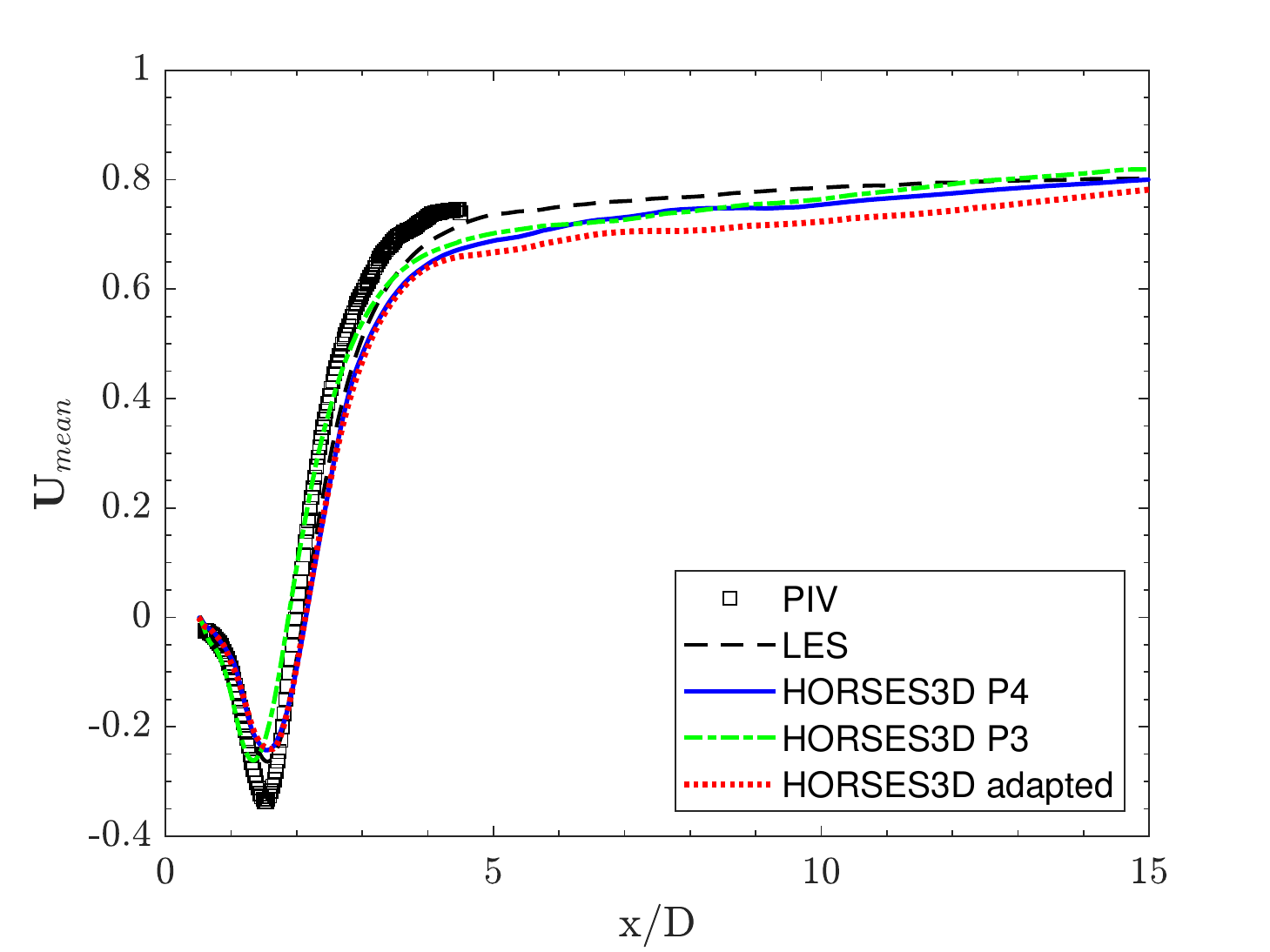}}
\subfloat[$U_{mean}$]{\label{fig:uy}\includegraphics[width=.45\linewidth]{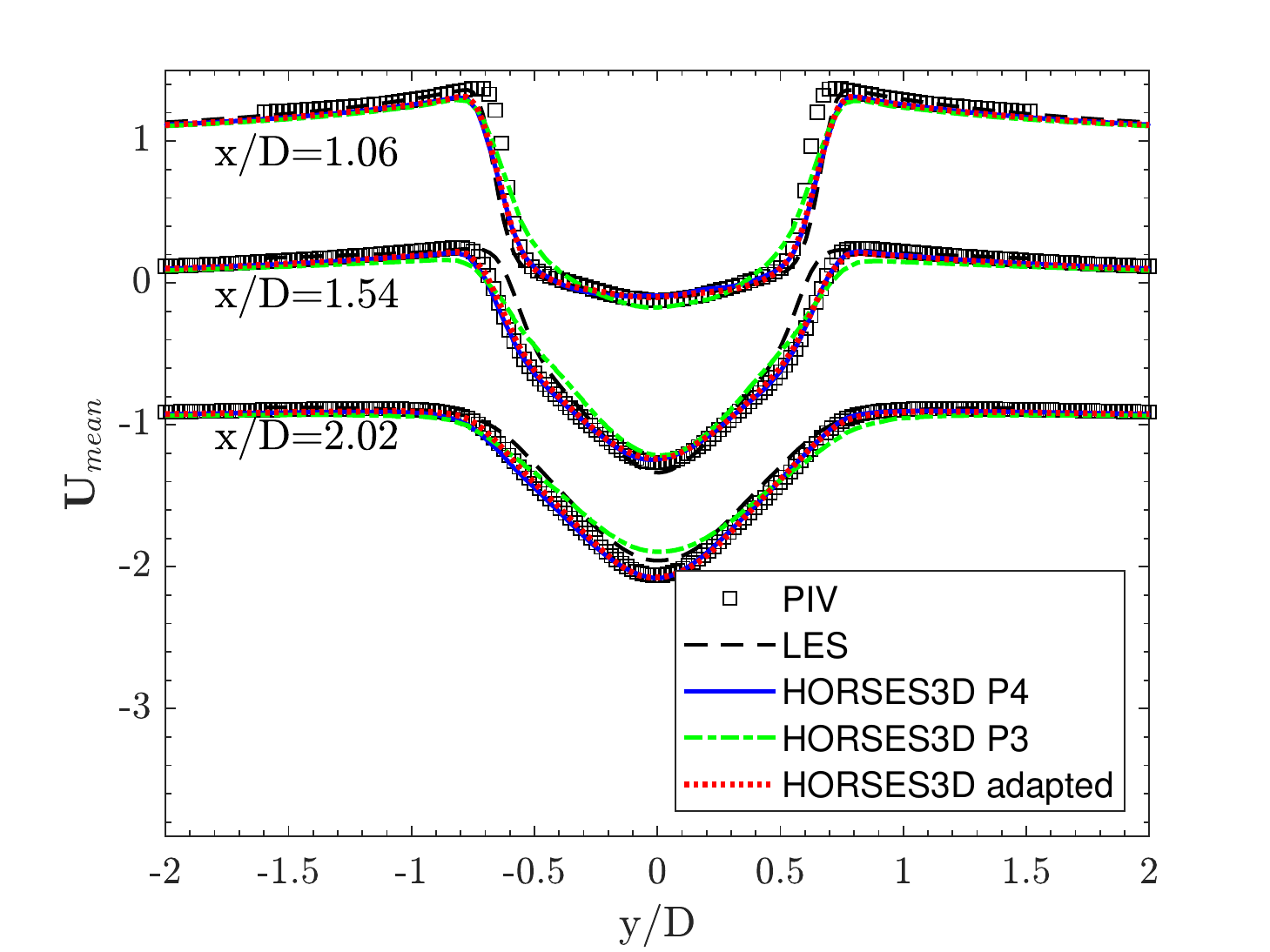}}\par
\subfloat[$V_{mean}$]{\label{fig:vy}\includegraphics[width=.45\linewidth]{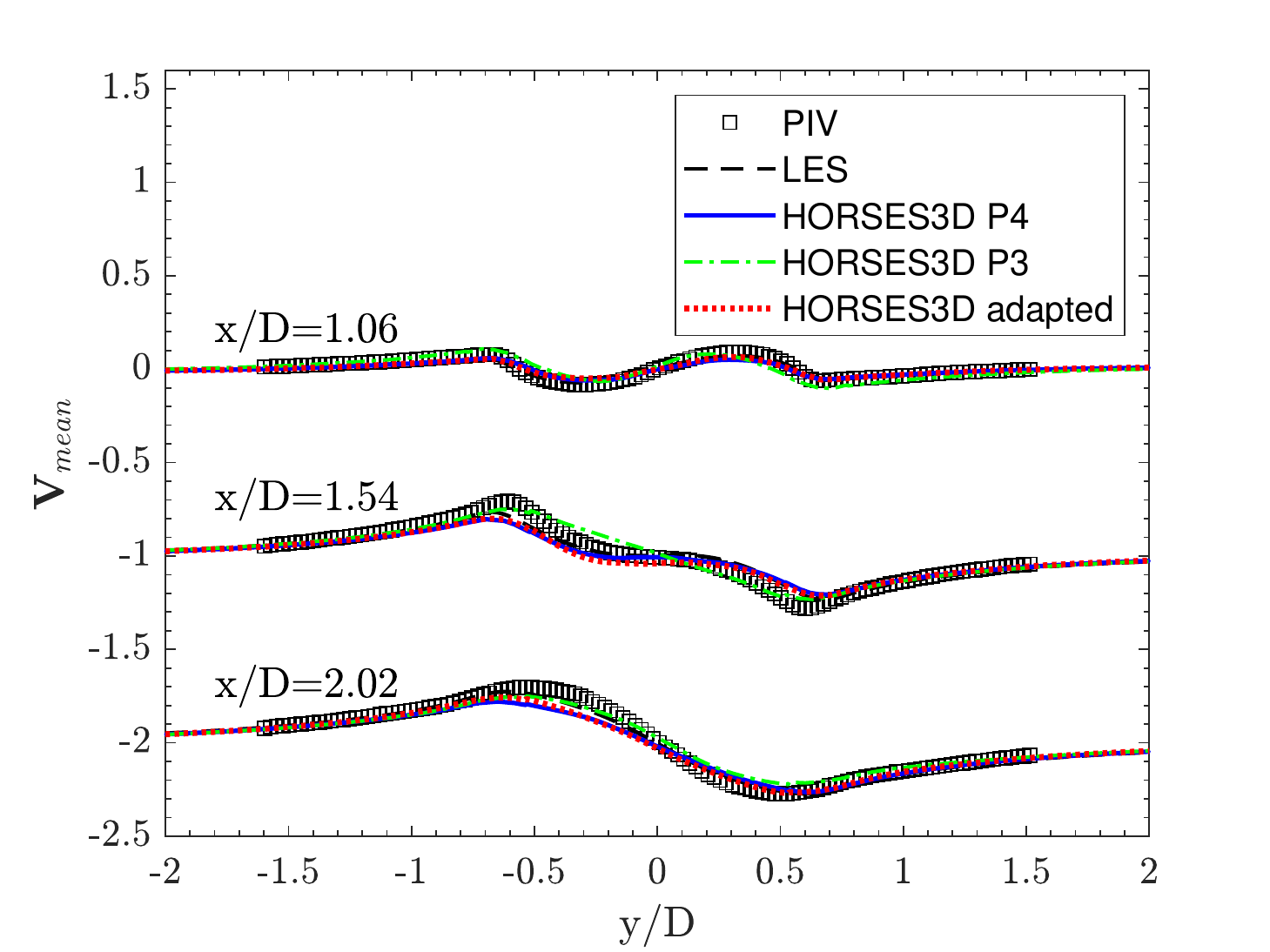}} 
\subfloat[$<v',v'>$]{\label{fig:vv}\includegraphics[width=.45\linewidth]{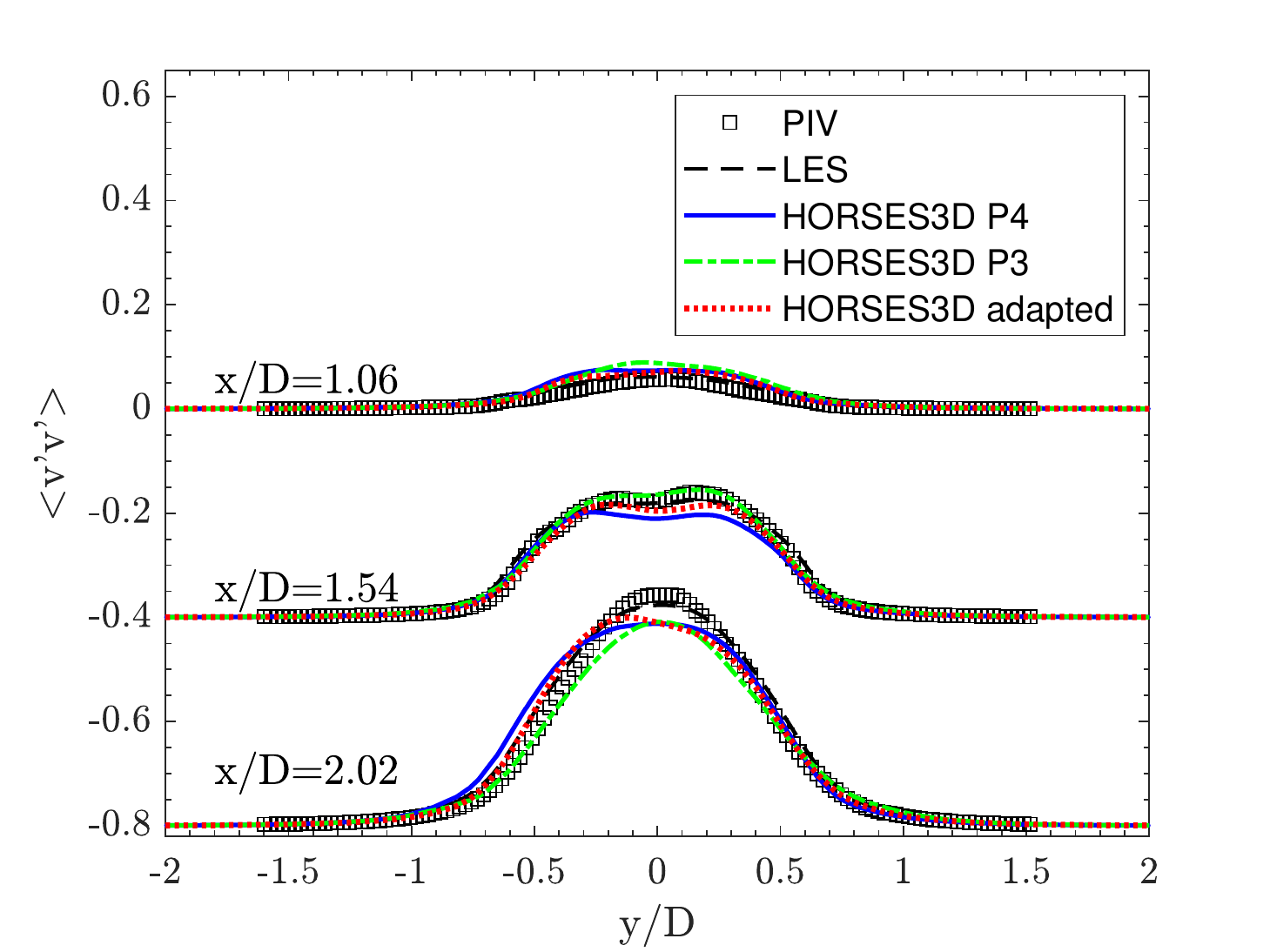}}
\caption{Statistical quantities of the wake for the unsteady turbulent case of a flow past a cylinder at Re=3900. Comparison of experimental (PIV) and high resolution numerical (LES) results from \cite{parnaudeau2008experimental} against uniform and adapted results from HORSES3D}
\label{fig:stats_cyl3900}
\end{figure}

 The clustering algorithm is able to successfully track the region of interest and through local $p$-refinement we attain a solution with $41\%$ less degrees of freedom, for a similar level of accuracy, compared to the uniform $P=4$ solution, as presented in table~\ref{tab:cyl3900_tab_dofs}. The adapted solution has only $16\%$ more degrees of freedom than the uniform $P=3$ solution but offers superior results as presented in table~\ref{tab:cyl3900_tab} and figure~\ref{fig:stats_cyl3900}. In terms of computational cost, as presented in table~\ref{tab:cyl3900_tab_dofs}, the adapted solution is $33$\% faster than the reference uniform $P=4$ solution for any given simulation interval. 

\begin{table}[!htb]
\centering
\caption{Comparison of the mean statistical quantities for the cylinder at Re=3900 results from HORSES3D and the literature }
\begin{tabular*}{0.8\textwidth}{c @{\extracolsep{\fill}} |ccc}
\hline
   & $\mathrm{DoFs}$ & reduction of DoFs &  reduction of comp. time  \\ \hline
Uniform P4     & 2.60M   & -     & -  \\ 
Uniform P3     & 1.33M   & 49\% & 50\%  \\ 
Cluster-Adapt P4-P2  & 1.55M   & 41\% & 33\%  \\ \hline
\end{tabular*}
\label{tab:cyl3900_tab_dofs}
\end{table}

The statistics for this test case are presented in figure~\ref{fig:stats_cyl3900}. The results of this work are compared against the experimental (PIV) and numerical results (LES) presented in \cite{parnaudeau2008experimental}. In figure~\ref{fig:uu} the results of the variance of the streamwise fluctuations for three different positions $x/D=[1.06,1.54,2.02]$ are presented. The adapted and uniform $P=4$ results are in good agreement with the PIV and LES data. It should be noted that in the position $x/D=1.06$ the two strong peaks, due to the transitional state of the shear layer \cite{parnaudeau2008experimental}, are lower for the uniform $P=3$ case which can be attributed to the lack of proper resolution \cite{parnaudeau2008experimental}. In figure~\ref{fig:uv} we present the covariance of the fluctuations. In this case, we also have good agreement between the reference data and the adapted and uniform $P=4$ solution. The uniform $P=3$ solution deviates, indicating that this level of resolution is not sufficient to capture the correct phenomena. The differences are less pronounced in the results presented in figures~\ref{fig:uy}, \ref{fig:vy} as we observe a good agreement for the mean streamwise and normal velocities between the reference results (LES, PIV) and the results of this work. The variance of the normal velocity fluctuations in figure~\ref{fig:vv} is in good agreement, although the uniform $P=3$, $P=4$ and clustering p-adaptation results of this work underestimate the peak along the position $x/D=2.02$. Lastly, the mean streamwise velocity along the centerline, presented in figure~\ref{fig:uc}, is close to the reference data but we observe that the point of minimum velocity is over-predicted in the results from HORSES3D.  

This second turbulent case shows that the clustering adaption can be very useful for turbulent unsteady cases. The GMM clustering is able to mark viscous and turbulent regions in the boundary layer and wake, where higher resolution is required and separate it from the outer inviscid region, where a lower resolution can be used.

\section{Conclusion}
The presented methodology has been successfully applied and resulted in accelerating the fluid flow simulation around a circular cylinder at Reynolds $40$ (steady laminar regime) and Re=3900 (unsteady turbulent regime) while maintaining the expected levels of accuracy. The local adaptation strategy decouples the viscous/turbulent region from the outer inviscid regions and through mesh refinement/coarsening it achieves comparable results to those with uniformly refined meshes.
For the laminar case, we have shown that the $p$-adapted solution with polynomial orders $(P_{cluster}=4, P_{inviscid}=[1,2,3]$ has the same accuracy as the homogeneous solution with a uniform polynomial order $(P = 4)$ while showing  67\%, 55\% and 35\% reduction in number of degrees of freedom and 32\%, 29\% and 19\% reduction of computational time, respectively.

Furthermore, we have shown that this method is applicable to more complex unsteady turbulent flows such as that past a circular cylinder at Re=3900. The adapted solution through the clustering method is similar to the benchmark uniform solution ($P=4$), as well as the experimental and numerical results reported in the literature, while having 41\% less degrees of freedom and 33\% reduced computational cost. This showcases that the algorithm tracks the regions of interest that need to be well resolved while allowing us to reduce the computational cost.

We conclude that the clustering-based adaptation is a useful tool within the feature-based methods and overcomes some of the classic drawbacks present in feature-based approaches. For example, the presented methodology does not require any threshold or iterative adaptation process to determine the refinement regions.

\section*{Acknowledgments}

Gerasimos Ntoukas and Esteban Ferrer acknowledge the financial support of the European Union’s Horizon 2020 research and innovation programme under the Marie Skłodowska-Curie grant agreement (MSCA ITN-EID-GA ASIMIA No 813605). 
Gonzalo Rubio acknowledges the funding received by the Grant SIMOPAIR (Project No. RTI2018-097075-B-I00) funded by \\ MCIN/AEI/10.13039/501100011033 and by ERDF A way of making Europe. 
Esteban Ferrer would like to thank the support of the Spanish Ministry \\ MCIN/AEI/10.13039/501100011033 and the European Union NextGenerationEU/PRTR for the grant "Europa Investigación 2020" EIN2020-112255, and also the Comunidad de Madrid through the call Research Grants for Young Investigators from the Universidad Politécnica de Madrid. 
Finally, all authors gratefully acknowledge the Universidad Politécnica de Madrid (www.upm.es) for providing computing resources on Magerit Supercomputer.

\section*{Data availability}
The data that support the findings of this study are available from the corresponding author upon reasonable request.


%
%
\bibliographystyle{unsrtnat}
\bibliography{biblio}

\end{document}